\documentclass[12pt]{article}

\usepackage[letterpaper, margin=1in]{geometry}
\usepackage{setspace}
\usepackage{lineno}
\usepackage{fancyhdr}
\usepackage{graphicx}
\usepackage{listings}
\usepackage{xcolor}
\usepackage{natbib}
\usepackage{url}
\usepackage{hyperref}
\usepackage{amsmath, amssymb}
\usepackage{algorithm}
\usepackage{algpseudocode}

\usepackage{pdfpages}

\begin{document}

{
\begin{spacing}{2}
\pagestyle{fancy}
\fancyhf{}
\fancyhead[L]{Talbot \& Bradburd -- ARGscape}
\fancyhead[R]{\thepage}
\renewcommand{\headrulewidth}{0pt}

\newcommand{\code}[1]{\lstinline!#1!}
\newcommand{\todo}{\textbf{\textcolor{red}{TODO}}}

\pagestyle{fancy}
\fancyhf{}
\fancyhead[L]{Talbot \& Bradburd -- ARGscape}
\fancyhead[R]{\thepage}
\renewcommand{\headrulewidth}{0pt}

\begin{center}
{\LARGE \textbf{\code{ARGscape}: A modular, interactive tool for manipulation of spatiotemporal ancestral recombination graphs}}

\vspace{1em}

Christopher A. Talbot$^{1,2,*}$ and Gideon S. Bradburd$^{1}$

\vspace{0.5em}

{\small
$^{1}$Department of Ecology and Evolutionary Biology, University of Michigan, 1105 North University Avenue, Ann Arbor, MI 48109, USA\\
$^{2}$Department of Computational Biology, Cornell University, 350 Tower Rd, Ithaca, NY 14853, USA
}

\vspace{0.5em}

{\small $^{*}$Corresponding author: \href{mailto:cat267@cornell.edu}{cat267@cornell.edu}}

\vspace{0.5em}

\end{center}

\vspace{1em}

\begin{abstract}

\noindent \textbf{Summary:} Ancestral recombination graphs (ARGs) encode the complete genealogical history of a population of recombining lineages. ARGs, and their succinct representation, tree sequences, are increasingly central to modern population genetics methods, yet building an intuition for ARGs remains challenging. This is particularly true when analyzing ancestry in a geographic context, as there is a critical lack of dedicated, interactive tools capable of visualizing ARGs as spatiotemporal objects. To address this gap, we introduce \code{ARGscape}, an interactive platform for simulating, analyzing, and visualizing ARGs across space and time. \code{ARGscape} provides a user-friendly graphical interface featuring dynamic 2- and 3-dimensional visualizations to explore ARGs through space and time, as well as a novel ``spatial diff" visualization for quantitative comparison of geographic inference methods. \code{ARGscape} is an innovative, unified framework that seamlessly integrates leading command-line, Python, and R-based tools for ARG simulation, manipulation, and use in spatiotemporal inference into both graphical and command-line interfaces. By integrating these various functionalities, \code{ARGscape} facilitates novel data exploration and hypothesis generation, while lowering the barrier to entry for spatiotemporal ARG analysis in both research and education use-cases.

\noindent \textbf{Availability and Implementation:} \code{ARGscape} is built with a Python FastAPI backend and a React/TypeScript frontend. It is freely available as a live demo at \url{https://www.argscape.com} [v0.3.0] and as a Python package on PyPI (\code{pip install argscape}) [v0.3.0]. The source code and documentation are available on GitHub at \url{https://github.com/chris-a-talbot/argscape}.

\noindent \textbf{Contact:} \href{mailto:cat267@cornell.edu}{cat267@cornell.edu}

\noindent \textbf{Supplementary information:} Supplementary Information is included at the bottom of this document.

\end{abstract}

\section{Introduction}

The study of genetic variation across time and space is fundamental to population genetics, providing insights into processes from disease outbreaks to species evolution \citep{beebeTracingTigerPopulation2013, breisteinGeographicVariationGene2022, rehmannSweepsSpaceLeveraging2025}. The record of these processes can be read from the Ancestral Recombination Graph (ARG), which traces the complete genetic ancestry, including recombination and coalescent events, for a set of samples \citep{rasmussenGenomeWideInferenceAncestral2014, lewanskiEraARGIntroduction2024b, brandtPromiseInferringUsing2024c, nielsenInferenceApplicationsAncestral2025a}. These structures can thereby provide unprecedented clarity in studies of genetic variation across continuous space and time, moving beyond discrete populations and timepoints and toward a representation of real-world evolutionary processes. Recent developments in ARG inference methods have made high-quality ARGs accessible for a wide range of datasets of varying size and quality \citep{kelleherInferringWholegenomeHistories2019, speidelMethodGenomewideGenealogy2019, hubiszMappingGeneFlow2020a, wohnsUnifiedGenealogyModern2022b, zhangBiobankscaleInferenceAncestral2023, dengRobustAccurateBayesian2024a}. In particular, the succinct tree sequence data structure, implemented by \code{tskit}, has made the storage and analysis of very large ARGs computationally tractable \citep{kelleherEfficientCoalescentSimulation2016a, ralphEfficientlySummarizingRelationships2020, wongGeneralEfficientRepresentation2024a}.

Consequently, ARGs are now central to a growing number of sophisticated inference methods \citep{hejaseDeepLearningApproachInference2022a, wohnsUnifiedGenealogyModern2022b, tagamiTstraitQuantitativeTrait2024, grundlerGeographicHistoryHuman2025a}. In particular, a growing area of research is the use of ARGs to infer the spatial history of ancestors of a population \citep{wohnsUnifiedGenealogyModern2022b, osmondEstimatingDispersalRates2024a, grundlerGeographicHistoryHuman2025a, derajePromiseChallengeSpatial2025a}. However, despite their theoretical power, a broadly accessible and unified framework for interpreting ARGs as spatiotemporal objects remains elusive. This gap hinders their adoption by the broader genomics community. While existing tools (such as \code{tsbrowse} and \code{tskit\_arg\_visualizer}, \citep{karthikeyanTsbrowseInteractiveBrowser2025, kitchensTskit_arg_visualizerInteractivePlotting2025}) provide powerful interfaces for inspecting the raw topology of and data stored in tree sequences and ARGs, they are not designed for the integrated spatial inference and visualization that is critical for many ecological and evolutionary questions.

To bridge this gap and enhance the accessibility of ARGs, we present \code{ARGscape}: an interactive platform for visualizing, exploring, and analyzing ARGs as spatiotemporal records of evolutionary history. \code{ARGscape} integrates simulation, analysis, and visualization into a single, intuitive workflow, making complex spatiotemporal analyses accessible to both researchers and students.

\section{System and Features}

\code{ARGscape} is a full-stack web application with a Python-based FastAPI backend and a React/TypeScript-based frontend, available as a publicly hosted website or a Python package for local deployment. Its workflow is designed to be modular and intuitive, guiding the user from data input to visualization and export. The backend design is intended to facilitate rapid integration of new algorithms and features, ensuring that developments in the field can be incorporated and made accessible quickly. The \code{ARGscape} Python package also features a command-line interface incorporating core functionality without the need to load the full web application.

\subsection{Data Input and Simulation}

Users can begin by either uploading existing tree sequences in \code{.trees} or compressed \code{.tsz} format or by simulating new ones. Custom spatial coordinates for nodes in the tree sequence can be supplied via \code{.csv} files. The simulation module uses \code{msprime} to generate tree sequences under a specified coalescent model \citep{kelleherEfficientCoalescentSimulation2016a, nelsonAccountingLongrangeCorrelations2020, baumdickerEfficientAncestryMutation2022}. For demonstrative purposes, \code{ARGscape} uses a multidimensional scaling algorithm on the genealogical distance matrix to produce ``toy" spatial coordinates for simulated samples, which can be generated on a 2-dimensional grid or projected onto real-world maps.

For command-line interface users, tree sequences can be loaded and stored in temporary memory using the \code{argscape\_load} command.

\subsection{Spatiotemporal Inference}

A core function of ARGscape is to provide a cohesive analytical environment that integrates multiple complex spatiotemporal inference tools, enabling direct comparison and hypothesis testing within a single interface. Once a tree sequence is loaded, users can perform:

\begin{itemize}
    \item \textbf{Temporal Inference:} \code{ARGscape} supports the full \code{tsdate} workflow, including ARG pre-processing and temporal inference \citep{speidelInferringPopulationHistories2021a}.
    \item \textbf{Spatial Inference:} Given sample locations in continuous 1- or 2-dimensional space, \code{ARGscape} enables rapid inference of ancestral node locations in geographic space. Geographic inference may be performed using any of the following models:
    \begin{itemize}
        \item \textbf{Wohns midpoint} \citep{wohnsUnifiedGenealogyModern2022b}
        \item \textbf{Gaia} [quadratic or linear algorithms] \citep{grundlerGeographicHistoryHuman2025a}
        \item \textbf{FastGaia} (see Supplementary Information, Section S2)
        \item \textbf{Sparg} \citep{derajePromiseChallengeSpatial2025a}
    \end{itemize}
\end{itemize}

These analyses are initiated with simple button clicks in the user interface, abstracting away the need for complex command-line syntax and separate data formatting for each approach. \code{ARGscape} also offers static and interactive command-line methods for manipulating tree sequences, including persistent storage of tree sequences and the full suite of spatiotemporal inference methods, using the \code{argscape\_infer} command.

\begin{figure}[p]
\centering
\includegraphics[width=\textwidth]{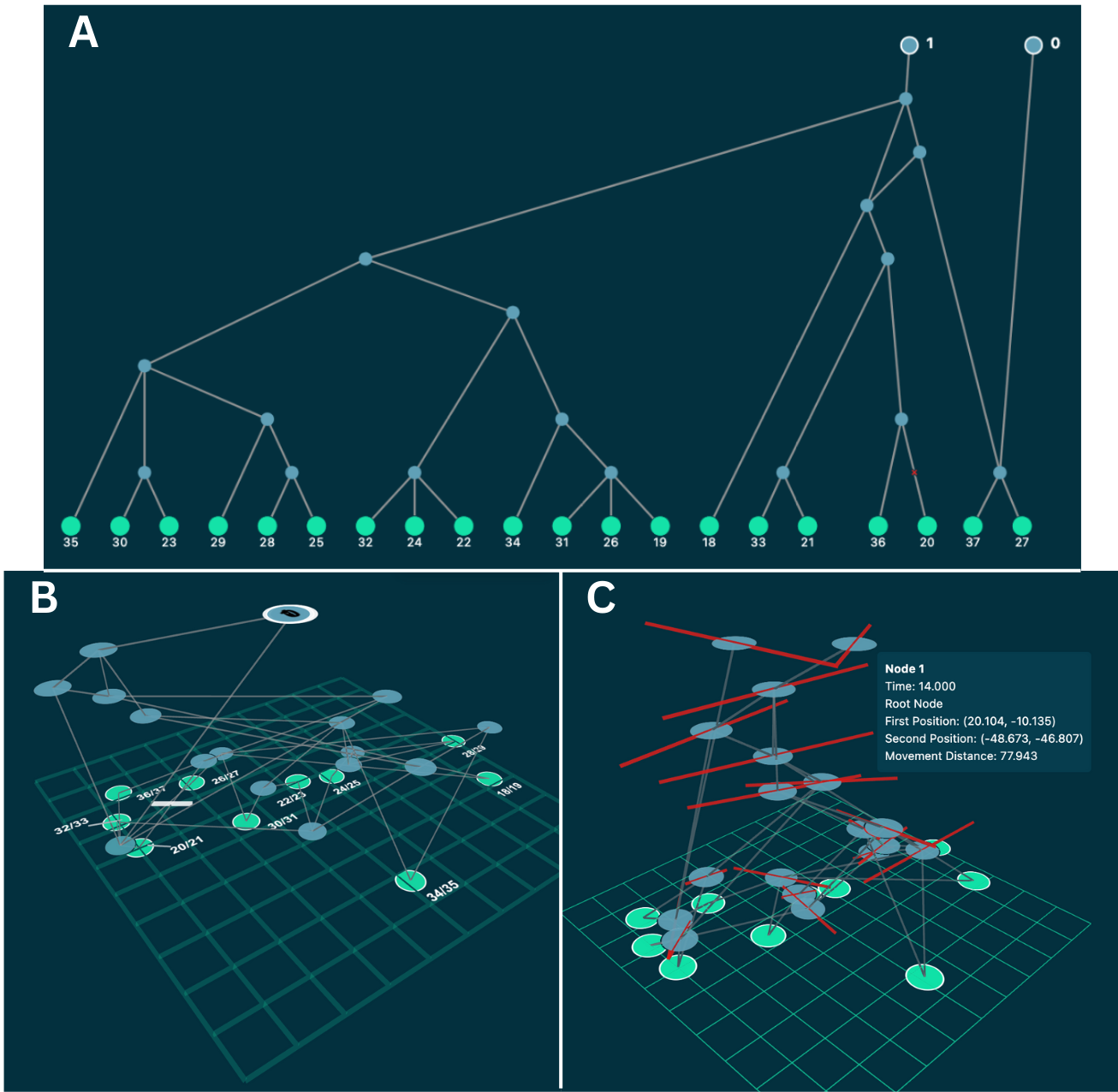}
\caption{\footnotesize Visualizations of an ancestral recombination graph (ARG) generated by a spatially explicit evolutionary simulation in \code{SLiM 5.0}. \textbf{A)} A 2D visualization, displaying the topology of the ARG. Our dagre-d3 mode generates this node ordering and layout. Green nodes represent samples, and blue nodes represent ancestral nodes. Blue nodes with a white outline are roots of the ARG. Branches with mutations along them are marked with a red ``x." Generations are evenly spaced along the y-axis. In \code{ARGscape}, node locations may be adjusted, node and edge labels toggled, and colors modified. Nodes may be selected to display parent or child subARGs. \textbf{B)} A 3D visualization of the same ARG with ground truth spatial data from the simulation. Green nodes represent samples, and blue nodes represent ancestral nodes. Blue nodes with a white outline are roots of the ARG. Branches with mutations along them are marked with white bars. Generations are evenly spaced along the z-axis. Sample nodes from the same individual are labeled together because they originate from the same location. In \code{ARGscape}, spacing may be adjusted, node and edge labels toggled, and colors modified. Nodes may be selected to display parent or child subARGs. \textbf{C)} A 3D ``spatial diff" visualization. Ancestral locations along the same ARG were re-inferred using the Gaia quadratic algorithm in \code{ARGscape}. Nodes are placed at the midpoint of the simulated and inferred locations, with red bars connecting the ground truth and inferred locations. This can be used to visualize the quantity and directionality of error in an inference method against ground truth; the increase in error moving back through time is clearly visible. Hovering over nodes reveals metadata, including their geographic locations in each version of the ARG and the distance between them.}
\label{fig:workflow}
\end{figure}

\subsection{Interactive Visualization}

\code{ARGscape} offers multiple visualization styles, each featuring extensive interactivity, customization, and analysis capabilities. Before visualizing with any method, the user may select a subset of the sample nodes, genomic regions, or temporal regions to visualize. This is particularly useful for ARGs that extend deep in time, for which visualizing only recent ancestry may be most informative. Across all visualizations, recombination nodes are combined into single nodes with multiple node IDs, which enhances intuitive visualization and increases rendering speed. \code{ARGscape} offers three primary visualization modes for exploring tree sequences:

\begin{enumerate}
    \item \textbf{2D Non-spatial View:} A force-directed graph rendered using \code{D3.js} displays the topological structure of the ARG. The force-directed graph visualization is fully customizable, featuring drag-and-drop node placement, toggleable node and edge labels, and a customizable color theme. To maximize interpretability, users can arrange sample nodes (tips) using multiple ordering algorithms specifically designed for tree sequences, including a novel ``consensus" method that aggregates node orders across multiple local trees. Internal nodes and edges are then placed using a force simulation to ensure adequate spacing for visual clarity. Additionally, the \code{dagre-d3} mode uses the \code{dagre-d3} React library to place all nodes and edges using a path-crossing minimization algorithm. Users can interactively explore the graph by clicking nodes to view sub-ARGs or trace ancestral lineages.
    
    \item \textbf{3D Spatial View:} A dynamic 3D plot rendered with \code{deck.gl} visualizes the ARG in geographic space, with time on the z-axis. The 3D spatial visualization is highly customizable, featuring toggleable node and edge labels, as well as a customizable color theme. The tool automatically detects the coordinate reference system (CRS) to display nodes on a unit grid or a world map. Users can upload custom shapefiles for bespoke geographic contexts. Interactive sliders for genomic and temporal windows enable powerful data exploration, allowing users to isolate and visualize how specific segments of ancestry have moved across space over time. Sample nodes that occur in identical locations and are $+/- 1$ node ID away from each other are assumed to represent haplotypes from the same individual, and are therefore combined into a single node for clarity. This specialized visualization mode opens doors for novel hypothesis generation and data exploration, including the visual assessment of coalescent rate through both time and space.
    
    \item \textbf{3D Spatial ``Diff" View:} A dynamic 3D plot rendered with \code{deck.gl} visualizes the difference in spatial locations on nodes in two ARGs with otherwise identical structure. Nodes are placed at the midpoint of their locations in each of two different ARGs, with a bar indicating the difference in locations between ARGs. This tool is one-of-a-kind and highly useful for comparing inferred ancestral locations with simulated ground truth, or comparing inferred locations between multiple different inference methods.
\end{enumerate}

\subsection{Accessibility and Data Export}

To cater to different user needs, \code{ARGscape} is available as both a public web application for easy access and testing, as well as a Python package for integration into local pipelines and resource-intensive use cases. Users can download any modified tree sequence (e.g., after spatial/temporal inference) in standard \code{.trees} or \code{.tsz} formats for further downstream analysis. Publication-quality images of any visualization can also be exported, facilitating integration into manuscripts, presentations, and educational materials.

\section{Conclusion}

\code{ARGscape} provides a much-needed, user-friendly platform that integrates the simulation, analysis, and spatiotemporal visualization of ancestral recombination graphs. By integrating powerful but complex inference tools in an intuitive interface and providing novel interactive visualizations, it significantly lowers the barrier to entry for researchers and students interested in exploring the full richness of ARGs. \code{ARGscape} simplifies the process of asking and answering questions about how ancestry unfolds across the genome over time and across geographic space. Future directions will focus on enhancing performance for BioBank-scale datasets and expanding the tool's analytical capabilities to visualize key demographic processes through time, including migration and admixture. We also aim to further develop its utility as an interactive learning platform for population genetics education by incorporating guided tutorials and visual-based lessons.

\section{Acknowledgments}

The authors thank the members of the Bradburd Lab at the University of Michigan, the Messer Lab at Cornell University, the tskit-dev team, and James Kitchens for their feedback and testing.

\section{Author contributions}

C.A.T. (Conceptualization [lead], Software [lead], Validation [lead], Writing -- original draft [lead], Writing -- review \& editing [equal]). G.S.B. (Supervision [lead], Funding acquisition [lead], Conceptualization [supporting], Writing -- original draft [supporting], Writing -- review \& editing [equal]).

\section{Funding}

This work was supported by the National Institutes of Health [grant R35-GM137919] to G.S.B.

\section{Data availability}

Source code and documentation are available at \url{https://github.com/chris-a-talbot/argscape}. Figure 1 uses a tree sequence generated by a spatially explicit simulation ran in SLiM 5.0 \citep{hallerSLiM5Ecoevolutionary2025}. The tree sequence and SLiM simulation script used to generate visualizations for Figure 1 are available on Zenodo under doi \url{10.5281/zenodo.17296543}.

\section{Conflict of interest}

None declared.

\bibliographystyle{plainnat}

\section*{Supplementary Information}

Supplementary Information is included at the bottom of this document.
\end{spacing}
}

\clearpage
\thispagestyle{empty}
\vspace*{\fill}
\begin{center}
    \rule{0.6\textwidth}{0.4pt}\\[1em]
    {\large \textbf{End of Primary Article}}\\[0.5em]
    {\it (Next: Supplementary Information)}\\[1em]
    \rule{0.6\textwidth}{0.4pt}
\end{center}
\vspace*{\fill}
\clearpage

{
\begin{spacing}{2}
\pagestyle{fancy}
\fancyhf{}
\fancyhead[L]{Talbot \& Bradburd -- ARGscape Supplementary Information}
\fancyhead[R]{\thepage}
\renewcommand{\headrulewidth}{0pt}

\newcommand{\code}[1]{\lstinline!#1!}

\lstset{
    basicstyle=\ttfamily\small,
    breaklines=true,
    frame=single,
    backgroundcolor=\color{gray!10},
    commentstyle=\color{green!50!black},
    keywordstyle=\color{blue},
    stringstyle=\color{red},
    showstringspaces=false
}

\pagestyle{fancy}
\fancyhf{}
\fancyhead[L]{Talbot \& Bradburd -- ARGscape Supplementary Information}
\fancyhead[R]{\thepage}
\renewcommand{\headrulewidth}{0pt}

\setcounter{section}{0}
\renewcommand{\thesection}{S\arabic{section}}
\renewcommand{\thesubsection}{S\arabic{section}.\arabic{subsection}}
\renewcommand{\thesubsubsection}{S\arabic{section}.\arabic{subsection}.\arabic{subsubsection}}

\begin{center}
{\LARGE \textbf{Supplementary Information}}

\vspace{0.5em}

{\large \textbf{\code{ARGscape}: A modular, interactive tool for manipulation of spatiotemporal ancestral recombination graphs}}

\vspace{1em}

Christopher A. Talbot$^{1,2,*}$ and Gideon S. Bradburd$^{1}$

\vspace{0.5em}

{\small
$^{1}$Department of Ecology and Evolutionary Biology, University of Michigan, Ann Arbor, MI, USA\\
$^{2}$Department of Computational Biology, Cornell University, Ithaca, NY, USA
}

\vspace{0.5em}

{\small $^{*}$Corresponding author: \href{mailto:cat267@cornell.edu}{cat267@cornell.edu}}

\end{center}

\vspace{1em}

\section{Installation \& Usage}

\subsection{Using the Hosted Web Application}

The latest production version of \code{ARGscape} is available at \url{https://www.argscape.com/}. The interactive interface guides users through the various modes of use and functionality, including file management, simulation, visualization, inference, and output downloading. Files stored on the hosted web server will be stored privately and securely for up to 24 hours. The web server storage is subject to being cleared, without notice, during updates. 

We recommend using the hosted application only for \textbf{educational} and/or \textbf{exploratory} purposes due to the potential for data loss and the limited computational capabilities of our web server. Users are limited to a small amount of computing power for simulation, inference, and visualization, which will be insufficient for most full-scale projects.

\subsection{Installing \code{ARGscape} Locally}

The latest production version of \code{ARGscape} is available as a Python package hosted on the Python Package Index (PyPI). To install, users must have Python version 3.8 or later. Then, from a terminal in which Python is available, users should run the following commands:

\code{> python -m pip install --upgrade pip}

\code{> python -m pip install argscape}

\noindent\code{ARGscape} will be installed in the Python environment.

\subsection{Running the Web Application Locally}

To run the web application hosted on a local machine, first see Supplemental Information 1.2 -- Installing \code{ARGscape} Locally. Once \code{ARGscape} is installed, the web application can be started from any terminal in which Python is available and \code{ARGscape} is installed using the following command:

\code{> argscape}

\noindent This command will launch the complete \code{ARGscape} web application in a browser window, hosted on your local machine. For additional options, including customizing the port \code{ARGscape} is hosted on and disabling certain features, run:

\code{> argscape --help}

\subsection{Using the Command-Line Tools}

To run the web application hosted on a local machine, first see Supplemental Information 1.2 -- Installing \code{ARGscape} Locally. Once \code{ARGscape} is installed, the command-line tools can be used from any terminal in which Python is available and \code{ARGscape} is installed. The available commands include:

\subsubsection{argscape}

The basic \code{argscape} command loads the web application in a browser window.

\subsubsection{argscape\_load}

The \code{argscape_load} command includes features for session file management, including loading tree sequences from files using

\code{> argscape_load load --file <filename>}

\noindent or, with sample and/or node locations, using 

\code{> argscape_load load-with-locations --file <filename> --sample-csv <filename> --node-csv <filename>}

To view the complete set of file management commands, run

\code{> argscape_load --help}

\subsubsection{argscape\_infer}

The \code{argscape_infer} command unifies features for running spatial and temporal inference methods on loaded tree sequences. A simple command-line interface is provided for selecting files, inference methods, and output locations, which can be accessed by running

\code{> argscape_infer}

To view the complete set of spatiotemporal inference commands, run

\code{> argscape_infer --help}

\section{FastGaia Algorithm}

\code{FastGaia} is an unpublished set of algorithms for inferring the geographic locations of ancestral nodes in a tree sequence given georeferenced samples. Drawing inspiration from \code{Gaia} and Wohns' midpoint approach \citep{wohnsUnifiedGenealogyModern2022b, grundlerGeographicHistoryHuman2025a}, this approach aims to utilize more of the information encoded in the ARG than a simple midpoint approach, while running faster than \code{Gaia} by implementing a parallelizable greedy algorithm approach. The ease of incorporating \code{FastGaia} within the \code{ARGscape} framework demonstrates the flexibility of integrating new methods into \code{ARGscape}'s modular framework. 

Like \code{Gaia}, \code{FastGaia} can operate in continuous space (no barriers to dispersal, Euclidean cost function) or discrete space (using a uniform transition cost between states or an input transition cost matrix). Note that \code{ARGscape} uses only the continuous-space version of the algorithm. The complete details of both \code{FastGaia} algorithms are provided below, along with the necessary notation.

\code{FastGaia} can be installed in Python environments running Python version 3.8 or later by running the command

\code{> python -m pip install fastgaia}

\subsection{\code{FastGaia} Algorithm Notation and Definitions}

\begin{itemize}
    \item $T$: Tree sequence with nodes $V$ and edges $E$
    \item $n = |V|$: Number of nodes
    \item $t(u)$: Time (age) of node $u$
    \item $S \subset V$: Set of sample nodes with known locations/states
    \item $E(u)$: Set of edges where $u$ is the parent
    \item $C(u) = \{v : (u, v) \in E\}$: Set of children of node $u$
    \item $P(u) = \{w : (w, u) \in E\}$: Set of parents of node $u$
    \item For edge $e = (u, v)$:
    \begin{itemize}
        \item $s(e)$: Genomic span (right - left coordinates)
        \item $b(e) = t(u) - t(v)$: Branch length (temporal distance)
    \end{itemize}
    \item $\mathcal{E}_u$: Set of valid edges from parent $u$ to children with known locations/states (used locally)
\item $\omega_e$: Weight assigned to edge $e$ (continuous inference)
\end{itemize}

\subsection{\code{FastGaia} Algorithm S1: Continuous Location Inference}

\textbf{Input:}
\begin{itemize}
    \item Tree sequence $T = (V, E)$
    \item Sample locations $\mathcal{L}_S = \{\ell_u \in \mathbb{R}^d : u \in S\}$
    \item Boolean flags: $w_{\text{span}}$, $w_{\text{branch}}$
\end{itemize}

\noindent\textbf{Output:} 
\begin{itemize}
    \item Inferred locations $\mathcal{L} = \{\ell_u \in \mathbb{R}^d : u \in V\}$
\end{itemize}

\vspace{1em}
\noindent\textbf{Algorithm S1: Continuous Location Inference}

\begin{algorithmic}[1]
\State Initialize $\ell_u \gets \text{NaN} \in \mathbb{R}^d$ for all $u \in V$
\State $\ell_u \gets \mathcal{L}_S(u)$ for all $u \in S$ \Comment{Assign known sample locations}
\State Partition nodes: $V_\tau = \{u \in V : t(u) = \tau\}$ for each unique time $\tau$
\State Sort times: $\mathcal{T} = \{\tau_1, \tau_2, \ldots, \tau_k\}$ where $\tau_1 < \tau_2 < \cdots < \tau_k$
\For{each time $\tau \in \mathcal{T}$} \Comment{Process from present to past}
    \For{each node $u \in V_\tau$} \Comment{Parallel processing possible}
        \If{$u \in S$}
            \State \textbf{continue} \Comment{Sample location already known}
        \EndIf
        \State $\mathcal{E}_u \gets \{e = (u, v) : v \in C(u) \text{ and } \ell_v \neq \text{NaN}\}$
        \If{$\mathcal{E}_u = \emptyset$}
            \State $\ell_u \gets \text{NaN}$ \Comment{No valid children}
            \State \textbf{continue}
        \EndIf
        \State \Comment{Compute weighted average location}
        \For{each edge $e = (u, v) \in \mathcal{E}_u$}
            \State $\omega_e \gets 1.0$
            \If{$w_{\text{span}} = \text{True}$}
                \State $\omega_e \gets s(e)$
            \EndIf
            \If{$w_{\text{branch}} = \text{True}$}
                \State $\omega_e \gets \omega_e \cdot \frac{1}{b(e)}$ \Comment{Inverse branch length}
            \EndIf
        \EndFor
        \State $W \gets \sum_{e \in \mathcal{E}_u} \omega_e$
        \If{$W = 0$}
            \State $\ell_u \gets \frac{1}{|\mathcal{E}_u|} \sum_{e=(u,v) \in \mathcal{E}_u} \ell_v$ \Comment{Equal weights fallback}
        \Else
            \State $\ell_u \gets \frac{1}{W} \sum_{e=(u,v) \in \mathcal{E}_u} \omega_e \cdot \ell_v$ \Comment{Weighted average location of children}
        \EndIf
    \EndFor
\EndFor
\State \Return $\mathcal{L} = \{\ell_u : u \in V\}$
\end{algorithmic}

\vspace{1em}
\noindent\textbf{Weight Formula:} For edge $e = (u, v)$ connecting parent $u$ to child $v$:
\[
\omega_e = 
\begin{cases}
s(e) \cdot \frac{1}{b(e)} & \text{if } w_{\text{span}} \land w_{\text{branch}} \\
s(e) & \text{if } w_{\text{span}} \land \neg w_{\text{branch}} \\
\frac{1}{b(e)} & \text{if } \neg w_{\text{span}} \land w_{\text{branch}} \\
1 & \text{if } \neg w_{\text{span}} \land \neg w_{\text{branch}}
\end{cases}
\]

\noindent\textbf{Location Update:}
\[
\ell_u = \frac{\sum_{v \in C(u)} \omega_{(u,v)} \cdot \ell_v}{\sum_{v \in C(u)} \omega_{(u,v)}}
\]

\subsection{\code{FastGaia} Algorithm S2: Discrete State Inference}

\textbf{Input:}
\begin{itemize}
    \item Tree sequence $T = (V, E)$
    \item Sample states $\mathcal{S}_S = \{\sigma_u \in \{1, 2, \ldots, m\} : u \in S\}$
    \item Cost matrix $M \in \mathbb{R}^{m \times m}$ (optional): $M_{ij}$ = cost of transition from state $i$ to state $j$
\end{itemize}

\noindent\textbf{Output:}
\begin{itemize}
    \item Inferred states $\mathcal{S} = \{\sigma_u \subseteq \{1, \ldots, m\} : u \in V\}$ (set-valued to account for ties)
\end{itemize}

\vspace{1em}
\noindent\textbf{Algorithm S2: Discrete State Inference}

\begin{algorithmic}[1]
\State Initialize $\sigma_u \gets \emptyset$ for all $u \in V$
\State $\sigma_u \gets \{\mathcal{S}_S(u)\}$ for all $u \in S$ \Comment{Assign known sample states}
\State Determine state space: $\Sigma = \{1, 2, \ldots, m\}$
\State Partition nodes: $V_\tau = \{u \in V : t(u) = \tau\}$ for each unique time $\tau$
\State Sort times: $\mathcal{T} = \{\tau_1, \tau_2, \ldots, \tau_k\}$ where $\tau_1 < \tau_2 < \cdots < \tau_k$
\For{each time $\tau \in \mathcal{T}$} \Comment{Process from leaves to root}
    \For{each node $u \in V_\tau$} \Comment{Parallel processing possible}
        \If{$u \in S$}
            \State \textbf{continue} \Comment{Sample state already known}
        \EndIf
        \State \Comment{Compute cost for each candidate state}
        \For{each candidate state $\alpha \in \Sigma$}
            \State $\mathcal{C}_{\text{child}} \gets 0$
            \For{each edge $e = (u, v)$ where $v \in C(u)$ and $\sigma_v \neq \emptyset$}
                \For{each state $\beta \in \sigma_v$}
                    \State $c \gets M_{\alpha,\beta}$ if $M$ provided, else $c \gets 1$
                    \State $\mathcal{C}_{\text{child}} \gets \mathcal{C}_{\text{child}} + c \cdot s(e) \cdot b(e)$
                \EndFor
            \EndFor
        \EndFor
        \State $c^* \gets \min_{\alpha \in \Sigma} \mathcal{C}_{\text{child}}$
        \State $\sigma_u \gets \{\alpha \in \Sigma : \mathcal{C}_{\text{child}} = c^*\}$ \Comment{All optimal states}
    \EndFor
\EndFor
\State \Return $\mathcal{S} = \{\sigma_u : u \in V\}$
\end{algorithmic}

\vspace{1em}
\noindent\textbf{Cost Formula:} For node $u$ and candidate state $\alpha$:
\[
\mathcal{C}(u, \alpha)
= \sum_{v \in C(u)} \sum_{\beta \in \sigma_v}
    c(\alpha, \beta) \cdot s_{uv} \cdot b_{uv}
\]

where the transition cost function is
\[
c(i, j) =
\begin{cases}
M_{ij}, & \text{if a cost matrix } M \text{ is provided} \\
1, & \text{otherwise (uniform cost)}
\end{cases}
\]

\noindent\textbf{State Assignment:}
\[
\sigma_u = \arg\min_{\alpha \in \Sigma} \mathcal{C}(u, \alpha)
\]

\noindent Note: $\sigma_u$ may include multiple states if there are ties for the minimum cost.

\section{2D Visualization Tip Ordering Algorithms}

While all 2D \code{ARGscape} visualizations are fully interactive and customizable, we also provide a diverse set of tree sequence-specific tip ordering algorithms designed to clarify complex, tangled graph layouts. The available tip ordering algorithms are detailed below. Ideal choice of tip ordering algorithm will vary on a case-by-case basis, with no clear general rules for which to choose in what scenario. However, especially for very complex graphs, \code{dagre-d3} mode will often produce the clearest graphs.

\subsection{Numeric}

The numeric tip ordering algorithm places sample nodes along the $x$-axis in order of increasing node ID. Ancestral nodes are placed using a force-directed simulation. This ordering will often result in complex and tangled visualizations.

\subsection{First Tree}

The first-tree tip ordering algorithm places sample nodes along the $x$-axis in the order of a minlex postorder traversal of the first local tree in the tree sequence. This order is generated automatically by \code{tskit} \citep{kelleherEfficientCoalescentSimulation2016a, ralphEfficientlySummarizingRelationships2020, wongGeneralEfficientRepresentation2024a}, and is the same ordering algorithm used by \code{tskit_arg_visualizer} \citep{kitchensTskit_arg_visualizerInteractivePlotting2025}. 

\subsection{Center Tree}

The center-tree tip ordering algorithm places sample nodes along the $x$-axis in the order of a minlex postorder traversal of the middle local tree in the tree sequence. If only one local tree is available, this is the same order as in first-tree. If two trees meet the ``center" criteria, the first is used.

\subsection{Consensus}

The consensus tip ordering algorithm orders sample nodes along the $x$-axis according to a majority vote by minlex postorder traversals across $K$ local trees in the tree sequence, where $K$ scales with number of local trees, and is bounded by $[1, 50]$. The $K$ local trees are selected from evenly spaced genomic positions across the tree sequence. If only one local tree is available, this is the same order as in first-tree or center-tree.

\subsection{Ancestral Path}

The ancestral path tip ordering algorithm aims to group sample nodes by shared ancestry and similar time to coalescence. It creates groups of samples that coalesced at similar times, then places groups with more recent shared ancestry closer to the center of the graph. This aims to minimize path crossings by putting groups with deeper ancestry -- and therefore longer edges -- towards the outside of the graph.

\subsection{Coalescence}

The coalescence tip ordering algorithm orders sample nodes along the $x$-axis in decreasing order of time to coalescence.

\subsection{Dagre-d3}

When \code{dagre-d3} mode is enabled, all nodes in the graph are placed by the \code{dagre-d3} React library using an edge-crossing minimization algorithm. The force-directed simulation is disabled when this mode is active. 

\bibliographystyle{plainnat}

\end{spacing}
}

\end{document}